Title: Emittance growth in linear induction accelerators

Author(s): Ekdahl, Carl A. Jr.

Intended for: Report

Issued: 2013-12-10

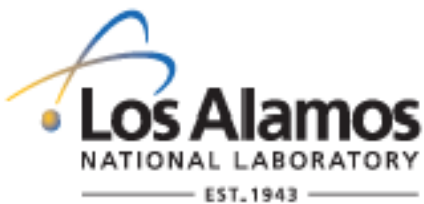

# Emittance growth in linear induction accelerators

Carl Ekdahl

*Los Alamos National Laboratory*

I. INTRODUCTION

Flash radiography of hydrodynamic experiments driven by high explosives is a well-known diagnostic technique in use at many laboratories [1]. Significant effort has been devoted to the development of the large electron accelerators needed for this [2]. At Los Alamos, the Dual-Axis Radiography for Hydrodynamic Testing (DARHT) facility provides multiple flash radiographs from different directions of an experiment. Two linear induction accelerators (LIAs) make the bremsstrahlung radiographic source spots for orthogonal views. The 2-kA, 20-MeV Axis-I LIA creates a single 60-ns radiography pulse. The 1.7-kA, 16.5-MeV Axis-II creates multiple radiography pulses by kicking them out of a 1600-ns long pulse from the LIA [3-5].

Beam emittance is the ultimate limitation on radiographic source spot size. In the absence of beam-target interaction effects, the spot size is directly proportional to the emittance. Since radiographic resolution is limited by the spot size, minimizing emittance enhances resolution of the radiographs. Therefore, investigation and mitigation of factors leading to high emittance beams would be a productive path to improved radiography.

Improvements in tuning the DARHT Axis-II LIA have reduced the beam motion during the four radiography pulses to less than 1-mm at the accelerator exit [5] and less than .04 mm at the final focus [2]. However, the issue of beam emittance has yet to be addressed. Although no measurements of the beam produced by the diode are available, detailed diode simulations with particle-in-cell (PIC) and particle-gun ray-trace codes predict a ~200-300 $\pi$-mm-mr normalized emittance. On the other hand, of the spot size at the final focus, and initial measurements of emittance in the downstream transport (DST) and imply an emittance > 900 $\pi$-mm-mr [3].

Two possibilities for this discrepancy are; our modeling of the diode is imperfect or there is emittance growth in the accelerator. With regard to our diode simulations, these assume a uniform emission from the hot dispenser cathode. Non-uniform emission is a known source of high emittance, as in DARHT Axis-I, where the high emittance from the cold explosive-emission cathode can be experimentally measured. Moreover, images of the Axis-II cathode show striking variations in luminosity, indicative of large temperature variation, and possible non-uniform emission. Moreover, small misalignment of the cathode assembly is another known source of high emittance [3]. The possibility of the diode causing the anomalously high emittance on Axis-II will be the subject of a future article. For the present article, I only investigate the possibility of emittance growth in the Axis-II LIA.

In addition to beam instabilities, there are at least four readily identifiable macroscopic mechanisms for emittance growth in an LIA.

- Spherical aberrations in the solenoidal focusing field.

- Helical beam trajectories with large gyro-radius.
- Strong dipole magnetic fields.
- Envelope oscillations.

These four mechanisms for emittance growth are amenable to investigation with particle-in-cell (PIC) simulations. A PIC code was used to simulate the beam in the DARHT Axis-II LIA, with particular attention to these mechanisms for emittance growth. The results of these simulations and their implications for DARHT are the subject of this article. Section II describes the simulation codes used, and Section III discusses the PIC simulations and results.

## II. SIMULATION CODES

Two simulation codes were used to explore the causes of emittance growth in a linear induction accelerator: the XTR envelope code and the LSP-slice PIC code. These are described in the next two subsections.

*Envelope Codes*

Design of tunes for the DARHT accelerators is accomplished using envelope codes. The two most frequently used are XTR and LAMDA. XTR was written by Paul Allison in the IDL language [6]. LAMDA was originally written by Tom Hughes and R. Clark [7]. In both of these codes the radius $a$ of a uniform density beam is calculated from an envelope equation [8]. In the DARHT accelerators the beam is born at the cathode with no kinetic angular momentum and a reverse polarity solenoid to cancel out the magnetic flux. Thus, the beam has no canonical angular momentum, and the envelope equation is

$$\frac{d^2 r}{dz^2} = -\frac{1}{\beta^2 \gamma}\frac{d\gamma}{dz}\frac{dr}{dz} - \frac{1}{2\beta^2 \gamma}\frac{d^2\gamma}{dz^2} r - k_\beta^2 r + \frac{K}{r} + \frac{\varepsilon^2}{r^3} \quad (1)$$

It can be shown that this same equation holds true for any axisymmetric distribution [9], so long as the radius of the equivalent uniform beam is related to the rms radius of the actual distribution by $r = \sqrt{2} R_{rms}$. Here, $\beta = v_e / c$, $\gamma = \sqrt{1 - 1/\beta^2}$, are the usual relativistic parameters, and the beam electron kinetic energy is $KE = (\gamma - 1) m_e c^2$. The betatron wavelength is

$$k_\beta = \frac{2\pi B_z}{\mu_0 I_A} \quad (2)$$

where $I_A = 17.08 \beta\gamma$ kA, and the generalized perveance is $K = 2 I_b / \beta^2 \gamma^2 I_A$. The emittance which appears in Eq. (1) is related to the normalized emittance by $\varepsilon = \varepsilon_n / \beta\gamma$, where

$$\varepsilon_n = 2\beta\gamma\sqrt{\langle r^2\rangle\left[\langle r'^2\rangle+\langle(v_\theta/\beta c)^2\rangle\right]-\langle rr'\rangle^2-\langle rv_\theta/\beta c\rangle^2} \qquad (3)$$

which is invariant through the accelerator under certain conditions.

The simple envelope equation in Eq. (1) is further improved in XTR as follows [10]. The energy dependence of the beam due to the gaps is approximated by a linear increase in $\gamma$ accompanied by a thin-einzel-lens focus. Between gaps $\gamma$ used in Eq. (1) is the value at the beam edge, which is space-charge depressed by $\Delta\Phi \approx 30 I_b(2\ln R_w/r)$, where $R_w$ is the radius of the beam pipe [8]. XTR also uses the magnetic field at the beam edge, including a first order approximation to to account for the flux excluded by a beam rigidly rotating in the magnetic field due to the invariance of canonical angular momentum [11].

The XTR code is used to design tunes for the DARHT Axis-II LIA. The envelope equation integrations in XTR are initiated at the exit of the diode, and initial conditions are provided by electron-gun design code simulations of the diode [3]. The beam envelope calculated by XTR for the tune used for most of 2013 is plotted in Figure 1. The initial envelope focusing is the result of tuning the six injector cells (z<500 cm) to prevent beam spill at any energy in the beam head, which slowly rises from zero to ~2.2 MeV at the diode exit in ~500 ns. The beam then rebounds through a focusing lattice designed to scrape off some of that off-energy beam head. This region is referred to as the beam-head cleanup zone, or BCUZ. The beam is then refocused into the main LIA.

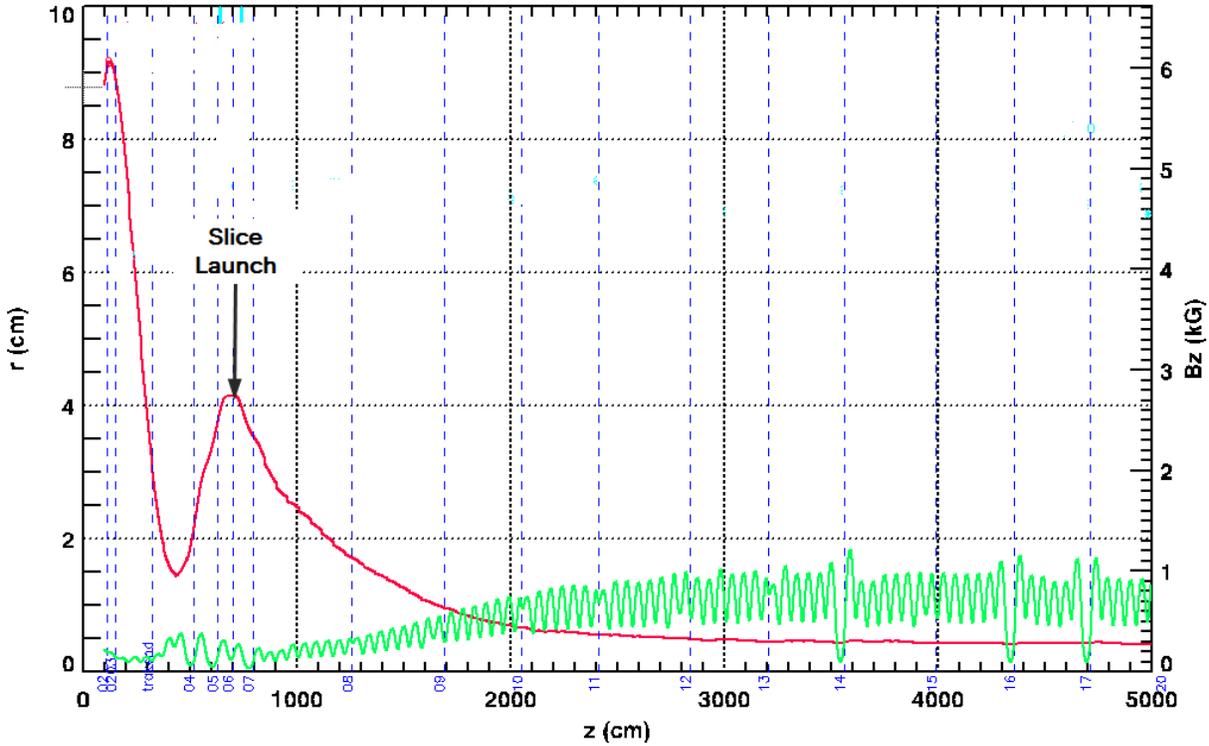

Figure 1: Envelope code simulation of beam transport through the injector cell block and into the main LIA. (Green) The solenoidal focusing magnetic field strength on axis (scale on right). (Red) The beam envelope radius (scale on left). (Blue, Dashed) Locations of beam position monitors (BPMs).

*Particle in Cell Codes*

The LSP-slice algorithm is a simplified PIC model for steady-state beam transport in which the paraxial approximation is assumed [12]. It is based on the Large Scale Plasma (LSP) PIC code [13]. A slice of beam particles located at an incident plane of constant z are initialized on a 2D transverse Cartesian ($x, y$) grid. The use of a Cartesian grid admits non-axisymmetric solutions, including beams that are off axis. Simulations were performed on a workstation with 32 processors. Multiprocessing reduced the time for a typical run from the more than 30 hours required for ealier single processor runs to less than 4 hours. For axisymmetric beams, one can use a faster version of the code based on a 1D cylindrical grid. Using all 32 processors, the typical 1D run completes in less than 4 minutes. The 1D version was often used when many runs were needed to investigate the effect of parametric variations. Excellent agreement between the 2D and 1D results have been obtained in comparison tests.

Initial electro- and magneto-static solutions are performed prior to the first particle push to establish the self-fields of the beam, including the diamagnetic field if the beam is rotating. After this initialization step, Maxwell's equations are solved on the transverse grid with

$\partial/\partial z = 0$, and then the particles are pushed by the full Lorentz equations. At each time-step the grid is assumed to be located at the axial center-of-mass of the slice particles $z(t)$, which is propagating in the $z$ direction.

The initial particle distribution of the slice is extracted from a full $x, y, z$ LSP simulation. The distribution is a uniform rigid rotor with additional random transverse velocity. The rotation is consistent with zero canonical angular momentum in the given solenoidal magnetic field at the launch position. The random transverse velocity is consistent with the specified emittance. Best agreement between LSP-slice and full LSP 2D simulations was obtained when the slice model is initiated at an envelope extreme, where the beam convergence is zero, so this condition was used for all simulations for this article. Also, for this article, 2D simulations used 70,688 particles, and 1D simulations used 4,000 particles.

External fields are input as functions of $z$, and are applied at the instantaneous axial center-of-mass location. External fields that are azimuthally symmetric (fields from solenoids and gaps) are input as on-axis values, and the off-axis components are calculated up to sixth order using a power series expansion based on the Maxwell equations [14]. In this way the nonlinearities of the accelerator optics are included in the slice simulations. The on axis magnetic field input was obtained from the XTR simulation shown in Fig. 1. Dipole steering fields are input as $x, y$ values that uniformly fill the solution space, an approximation that is obviously best for a beam near the axis. These dipole fields were also obtained from XTR, which calculates them on axis from measured cell misalignments and steering dipole excitation currents.

Although the envelope equation only deals with axisymmetric beams centered on axis, the concept of beam emittance is much more general, and it can be calculated for non-axisymmetric distributions in LSP-slice simulations. Consider a non-rotating beam with normalized distribution $\rho(x, x')$ in the $(x, x')$ plane of phase space. The position of the centroid of this distribution is at

$$\langle x \rangle = \iint x \rho(x, x) dx dx'; \quad \langle x' \rangle = \iint x' \rho(x, x) dx dx' \qquad (4)$$

Now consider the $2 \times 2$ matrix with elements given by

$$\begin{aligned}
\sigma_{xx} &= \iint (x - \langle x \rangle)^2 \rho(x, x') dx dx' = \langle x^2 \rangle - \langle x \rangle^2 \\
\sigma_{x'x'} &= \iint (x' - \langle x' \rangle)^2 \rho(x, x') dx dx' = \langle x'^2 \rangle - \langle x' \rangle^2 \\
\sigma_{x'x} &= \sigma_{xx'} = \iint (x - \langle x \rangle)(x' - \langle x' \rangle) \rho(x, x') dx dx' = \langle xx' \rangle - \langle x \rangle \langle x' \rangle
\end{aligned} \qquad (5)$$

This sigma matrix for the beam is

$$\sigma_x = \begin{pmatrix} \sigma_{xx} & \sigma_{xx'} \\ \sigma_{x'x} & \sigma_{x'x'} \end{pmatrix} \qquad (6)$$

This matrix is directly related to the area occupied by the beam in the $x, x'$ cut through phase space; $A = 4\pi\sqrt{\det\sigma}$. Since the emittance is related to the phase are by $\varepsilon_{rms} = A/\pi$, it follows that the rms emittance in the $x, x'$ cut through phase space is $\varepsilon_{xrms} = 4\sqrt{\det\sigma_x}$. For rotating beams with coupled coordinates and momenta, this can be easily generalized to $\varepsilon_{rms} = 4\sqrt{\det\sigma}$, where $\sigma$ is the $4\times 4$ sigma matrix formed from moments like Eq. (4) and Eq. (5) permuted through all transverse coordinates $x, x', y, y'$. This is how the emittance is calculated in LSP [15]. The area of the phase ellipse in the $x, x'$ plane is simply related by the sigma matrix to the rms emittance: $A_x = \beta\gamma\pi\varepsilon_{xrms} = \pi\varepsilon_{nx}$ $A_x = \beta\gamma\pi\varepsilon_{xrms} = \pi\varepsilon_{nx}$, so to the extent that phase space is conserved by Liouville's theorem, so is the normalized emittance. In the LSP-slice code, the normalized emittance is calculated as $\varepsilon_n = 4\beta\gamma\sqrt{\det\sigma}$. This reduces to Eq. (3) for a centered axisymmetric beam, which may be rotating.

III. RESULTS

I based these simulations on the magnetic tune used on the Axis-II LIA throughout 2013 (Fig. 1). All LSP-Slice runs were initiated just downstream of the BCUZ as shown in Fig. 1. At this location, the beam envelope is a maximum with no convergence, as calculated by envelope and PIC simulations beginning at the diode. Launching the LSP-slice code at such a location has been shown to give the best agreement with full PIC simulations. The beam size is much smaller here than at the diode, which improves the approximations in the codes, and reduces differences between them. This also eliminates errors in LSP-slice that would result from space-charge errors due to different tube size in injector-BCUZ-main LIA. Tube size is constant from here, and higher energy reduces some other differences. Earlier runs suggested < 20% growth of normalized emittance $\varepsilon_n$ at this point.

Magnetic fields on axis that were input to LSP-Slice were obtained from XTR, which includes an algorithm for calculating the dipole fields resulting from a table of the measured cell misalignments, as well as the axial field from the focusing solenoids [5]. External electric fields were derived from the locations, geometry, and voltage of the gaps in the XTR simulation in Fig.1.

*Baseline Simulation.*

For the baseline case corresponding to the tune of Fig 1, a beam concentric with the axis was injected at the launch point shown in Fig. 1. This launch point was chosen because it is an inflection of the beam envelope, which has been shown to produce the best results with the LSP-slice code. Beam parameters at the launch point were from the envelope code, and are listed in Table I. Figure 2 shows a comparison of results from the XTR envelope code and the LSP-slice code launched with these parameters. Since the LSP-Slice output for beam size is the rms radius, this was converted to the equivalent envelope radius vis $r = \sqrt{2}R_{rms}$ whenever necessary.

Table I. Initial beam parameters for LSP-slice code simulations.

| Launch Position | Axial Magnetic Field | Initial Envelope Radius | Initial Envelope Convergence | Normalized Emittance | Beam Current | Wall Potential |
|---|---|---|---|---|---|---|
| $z_0$ | $B_{z0}$ | $r_0$ | $r_0'$ | $\varepsilon_n$ | $I_b$ | $\varphi_{Wall}$ |
| cm | Gauss | cm | mradian | $\pi$-cm-rad | kA | MV |
| 703.11 | 160.6 | 4.1502 | 0.0 | 0.0206 | 1.68 | 3.29 |

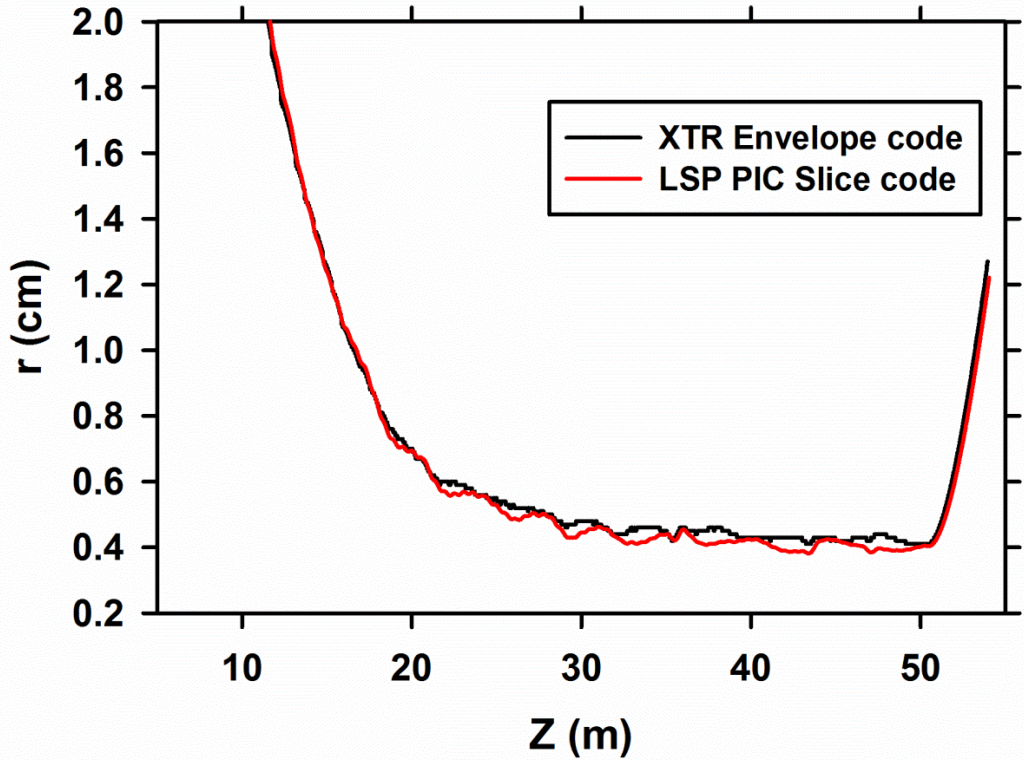

Figure 2: Beam envelope calculated with XTR and 2D LSP-slice for the tune shown in Fig. 1..

When comparing results of the LSP-slice code to those of the XTR envelope code, it is well to remember some of the differences in the physics and approximations thereof.

- Space charge depression is approximated in XTR. It is explicit in LSP-slice.
- XTR approximates space charge depression for variable beam tube size, but tube size in LSP-slice is constant.
- Beam diamagnetic depression of solenoidal guide field is approximated in XTR. It is explicit in LSP-slice.
- Different approximations are used by the two codes for the accelerating fields in gap regions.
- XTR uses a thin lens approximation for gap focusing. It is explicit in LSP-slice.
- XTR has corrections for image forces at gaps. This is missing in the LSP-slice model.

- Angular momentum: A rigid rotor with exactly zero canonical angular momentum is assumed in XTR. The rigid rotor in LSP-slice based on $B_z$, not $rA_\theta$, so it is less accurate for large beams in regions where $B_z$ varies with $r$ (fringe fields).

The initial emittance, $\varepsilon_n = 0.0206\ \pi$-cm-radian, is based on earlier PIC-slice code runs beginning at the diode exit. The resulting emittance through the accelerator from the LSP-slice simulation is shown in Fig. 3. The absence of growth for this tune is at odds with experimental estimates of emittance obtained from focusing magnet scans. These show emittance in excess of $0.07\ \pi$-cm-radian, so it is important to understand the physical root of this discrepancy. Thus, I examined the obvious causes in turn.

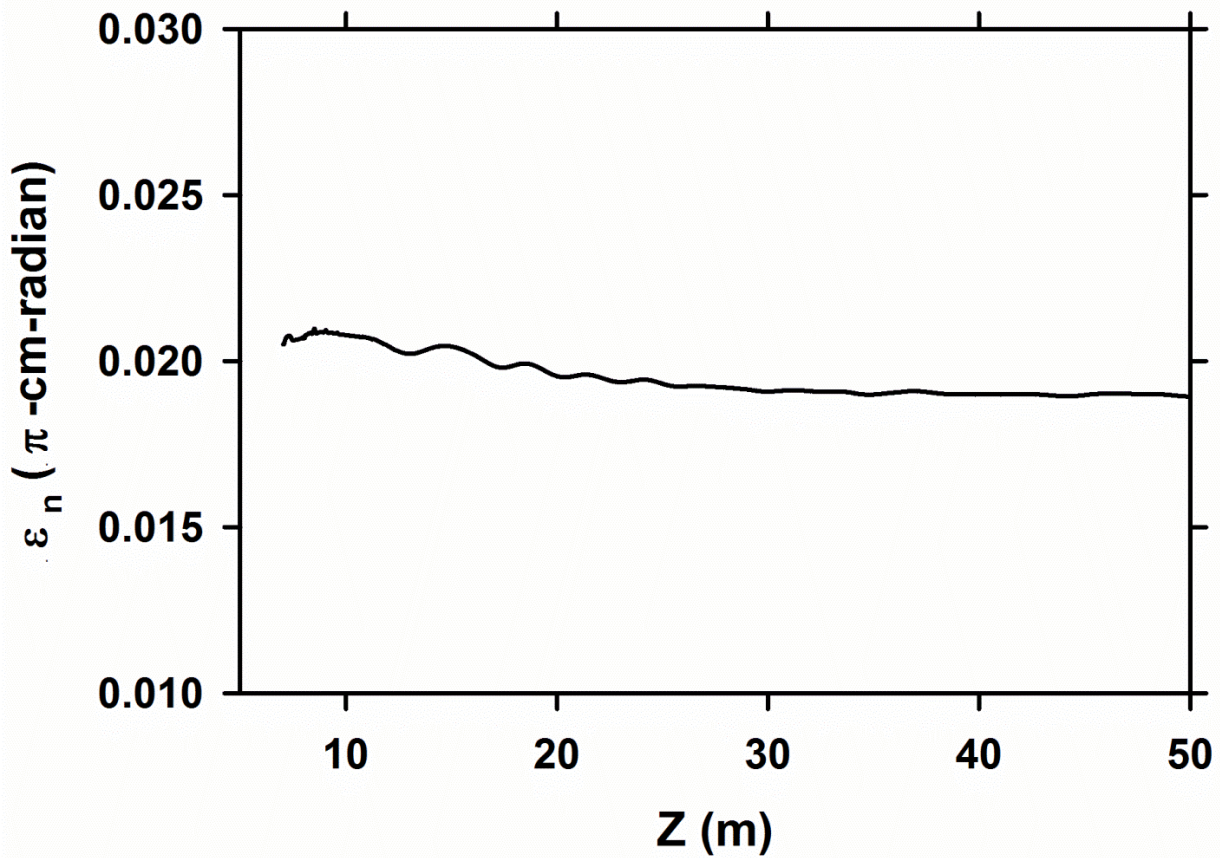

Figure 3: Normalized emittance $\varepsilon_n$ calculated by LSP-slice for the tune shown in Fig. 1. (The envelope calculated by LSP-slice is shown in Fig. 2.)

*Edge Focusing*

A well-known contributor to emittance growth in solenoidal focusing systems which spherical aberrations [16, 17], which over-focuses the edge of the beam, producing hollow beam profiles. The field expansions in these simulations include these aberrations, but even though the cumulative spherical aberration is large in our LIA there is apparently no emittance growth due to this effect in our baseline simulation (Fig. 3). This is probably because the beam is rapidly focused to a size much smaller than the bore of the solenoids, thereby minimizing the effect of the aberrations.

*Helical Trajectories*

Off center beams can have large helical trajectories in the solenoidal transport field. If the gyro-radius is too large, the beam distribution becomes distorted and the emittance increases. To demonstrate this effect, I initialized the beam for LSP-slice at different values of $x_0$, adjusting $y'_0$ to produce a helical trajectory of the beam centroid that encircled the axis. Fig. 4 shows the centroid trajectory calculated by LSP-slice for an initial offset of 1 cm, showing how the gyroradius is rapidly focused by the first few solenoids. Figure 5 shows an end on view of the helical trajectory, and also shows the approximate 2:1 focusing to equilibrium in the LIA. Fig. 6 compares the beam distributions at the exit for several different initial offsets, showing the severe distortion resulting from the largest gyro-radii. Figure 7 shows the emittance growth through the accelerator resulting from the beam distortion for each of these offsets. Obviously, the largest helical trajectories are very dangerous, because of the severe beam distortion and disastrous emittance growth. The results are summarized in Table II. For reference, as measured by our beam position monitors (BPMs) the beam in Axis-II is within 0.5-cm of the axis through most of the LIA, so emittance growth of more than 10% from this mechanism is not expected.

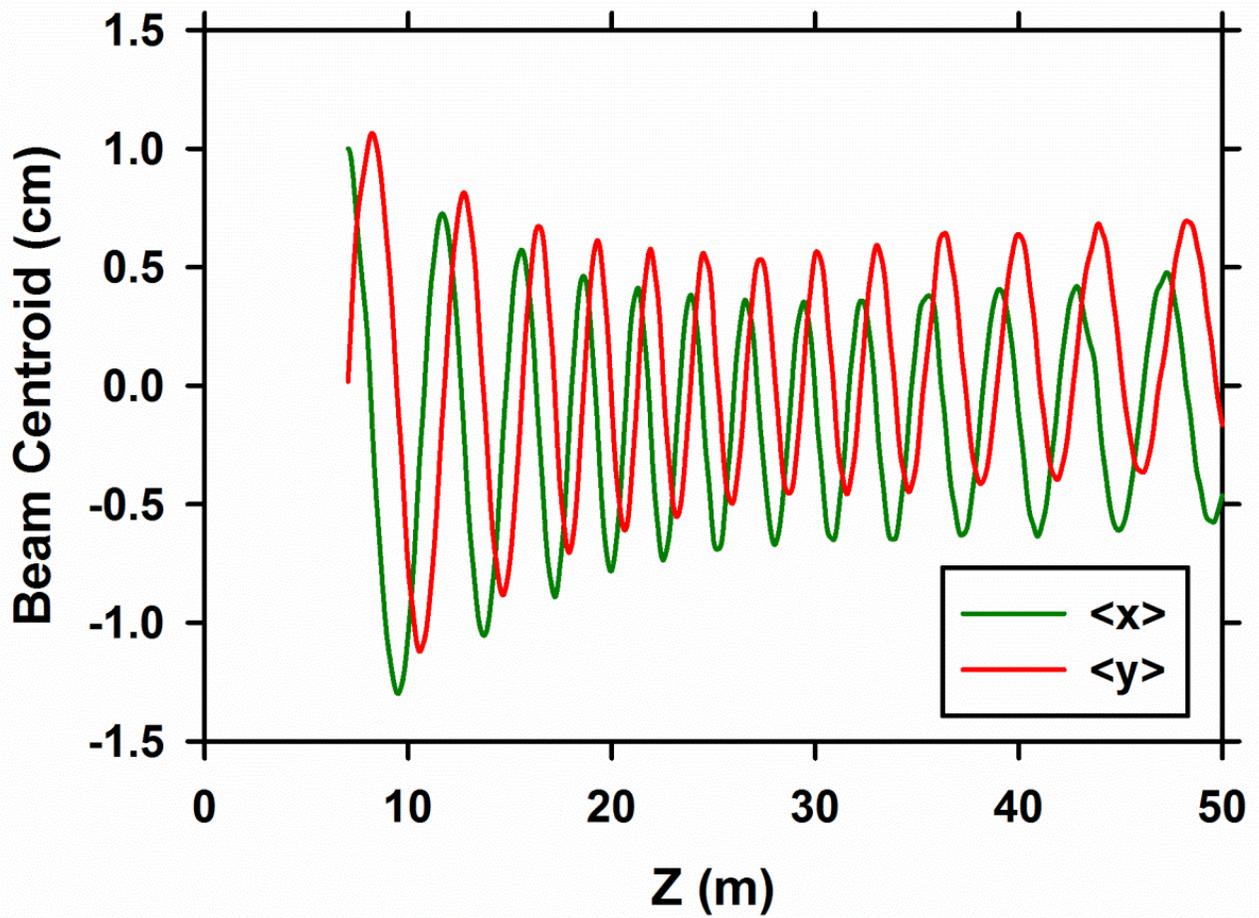

Figure 4: Trajectory of beam centroid calculated by LSP-slice. The beam was initially injected at $(x_0, y_0) = (1,0)$ cm. (Green) X coordinate of beam centroid. (Red) Y coordinate of beam centroid.

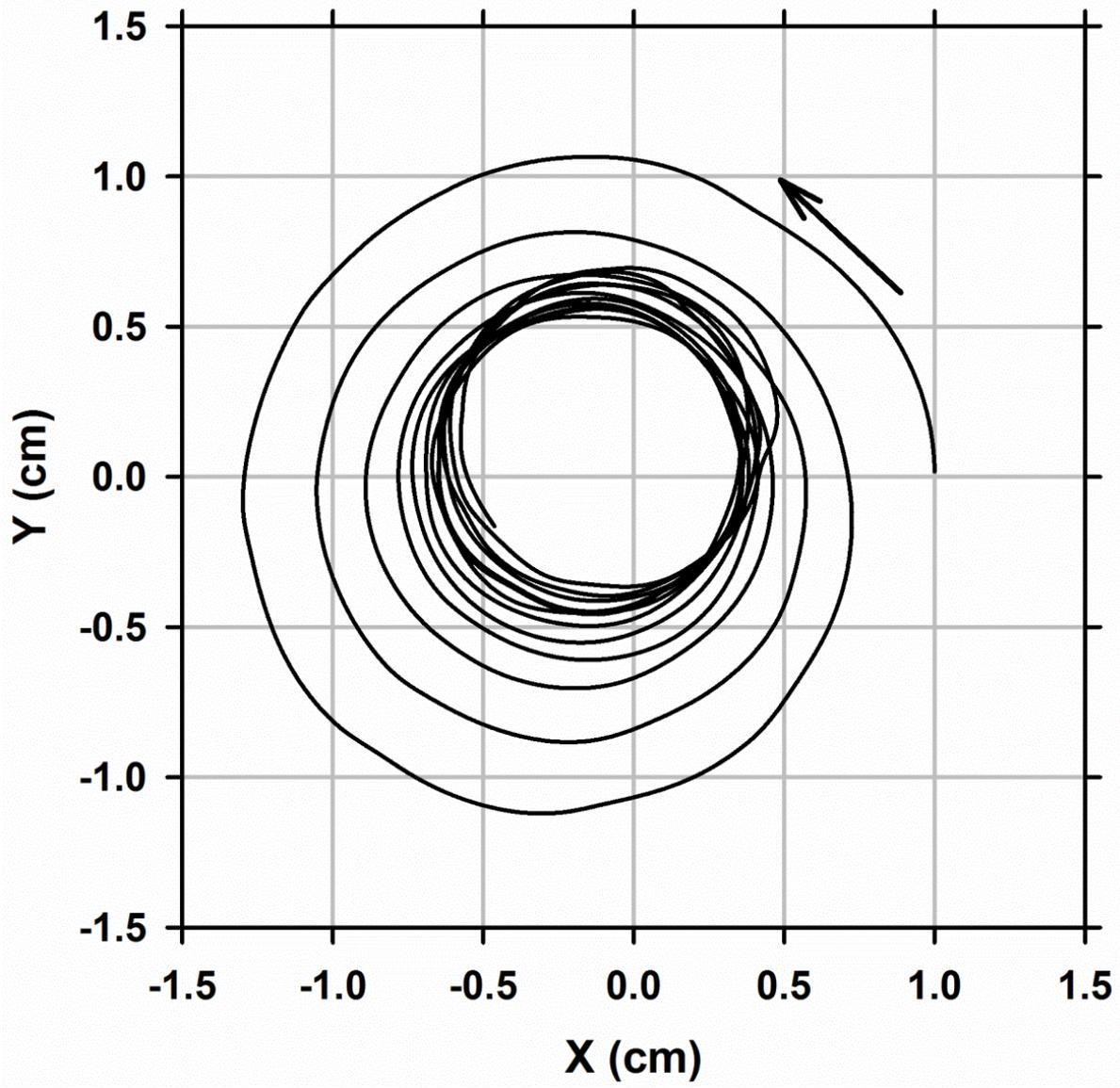

Figure 5: End on view of trajectory shown in Fig. 4. The direction of motion is indicated by the arrow.

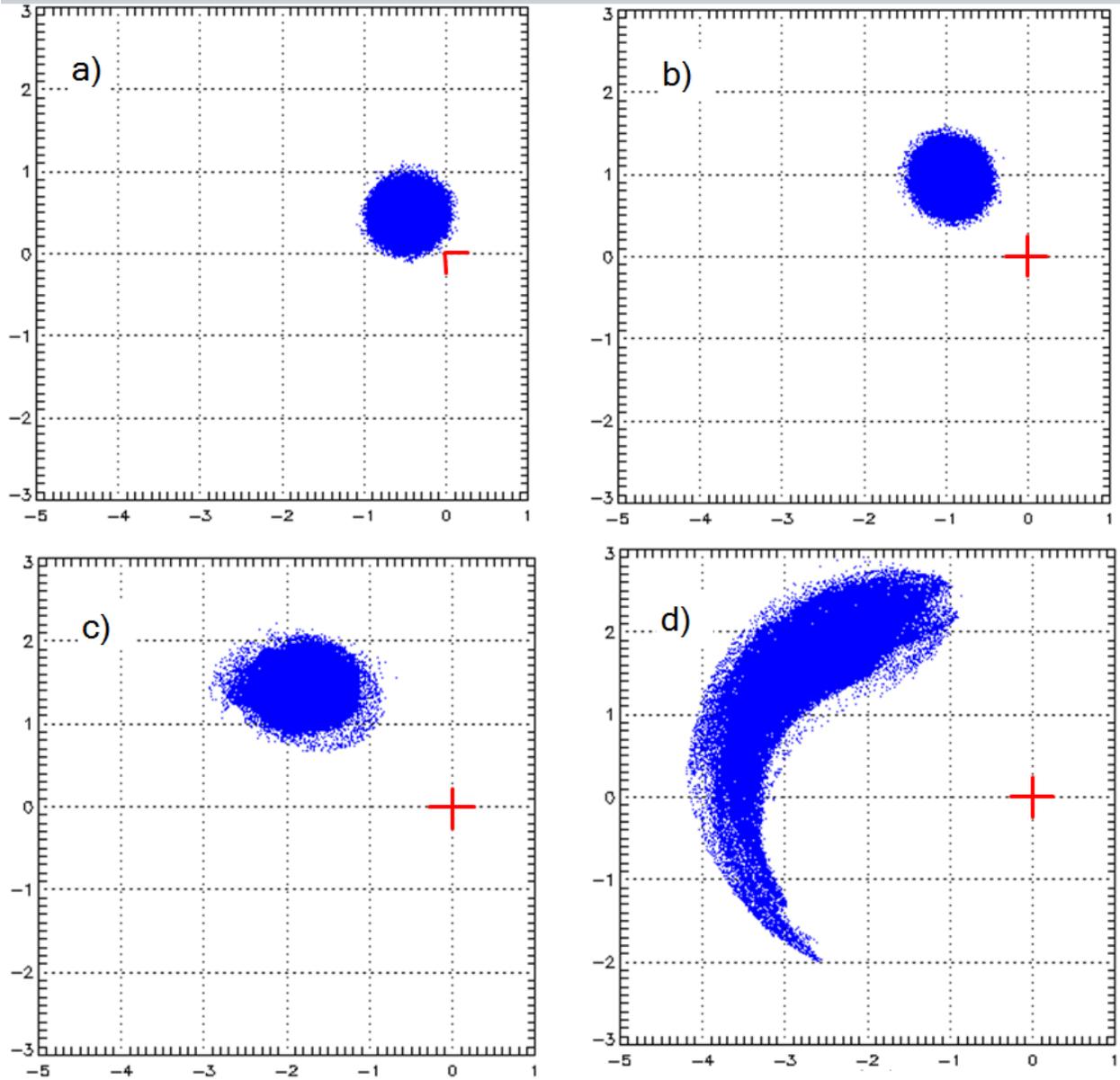

Figure 6: Distributions for axis-encircling beams at the exit of the LIA. The scales and grids are in 1-cm increments. In all cases the red cross marks the axis of the LIA.
a) $(x_0, y_0) = (1,0)$ cm. b) $(x_0, y_0) = (2,0)$ cm. c) $(x_0, y_0) = (4,0)$ cm. d) $(x_0, y_0) = (6,0)$ cm.

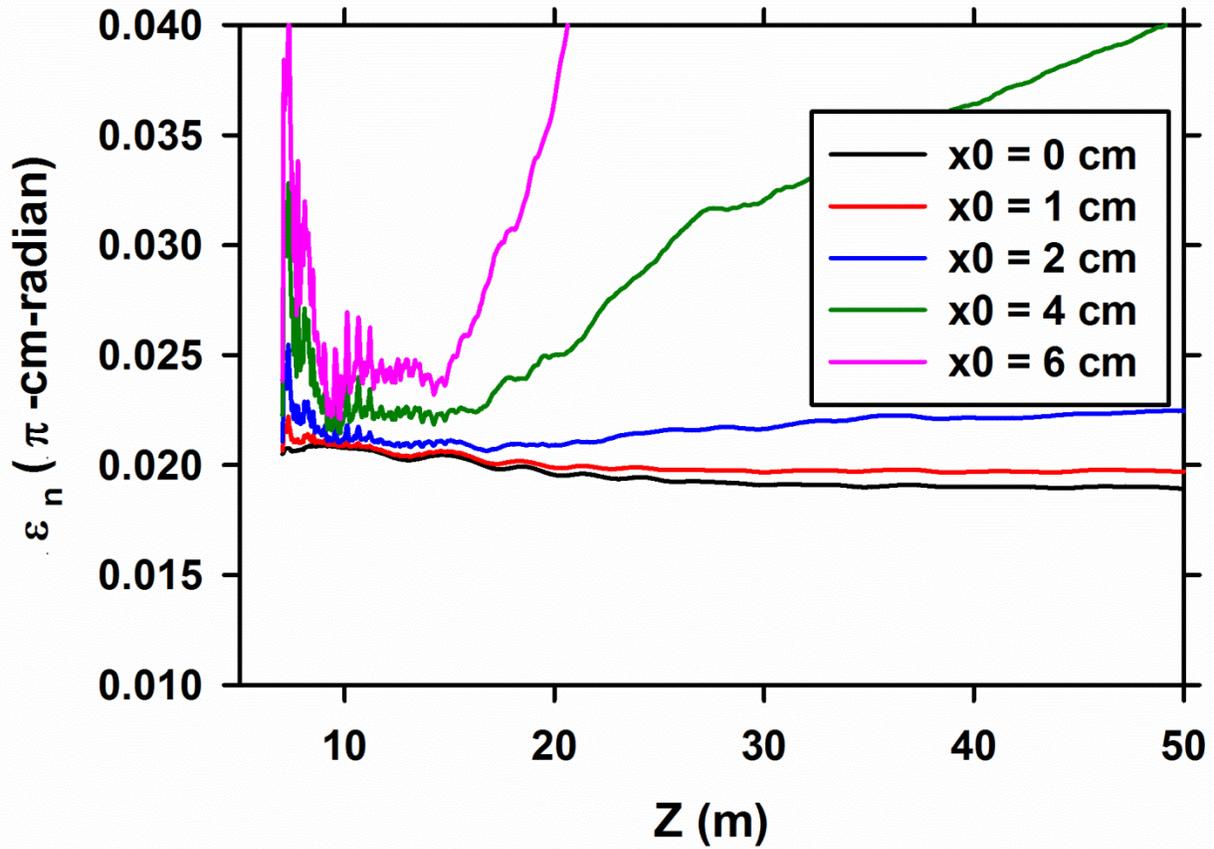

Figure 7: Normalized emittance growth resulting from axis-encircling beams.

Table II: Emittance growth from LSP-slice simulations of beams with large helical trajectories.

| $x_0$ cm | $y_0'$ mr | Initial gyroradius cm | Equilibrium gyroradius cm | Final emittance π-cm-radian | % Increase over baseline |
|---|---|---|---|---|---|
| 0.0 | 0.000 | 0.0 | 0.0 | 0.0190 | 0 |
| 1.0 | 0.015 | 1.0 | ~0.5 | 0.0197 | 4 |
| 2.0 | 0.030 | 2.0 | ~1.0 | 0.0225 | 18 |
| 4.0 | 0.055 | 4.0 | ~1.7 | 0.0399 | 110 |
| 6.0 | 0.080 | 6.0 | ~2.7 | 0.1884 | 892 |

*Magnetic Dipoles*

Another easily identified mechanism for emittance growth is magnetic dipole steering. This is be linked to helical motion, since steering with dipoles can cause or correct helical motion of off-center beams. One source of magnetic dipoles in the DARHT Axis-II LIA is cell-to-cell misalignment. Although substantial efforts were made to ensure alignment of the magnetic axis, small misalignments exist (~0.0 25-mm rms offset, and ~0.3-mr rms tilt). Beam energy variations coupling with such misalignments is the source of the "corkscrew" motion [18] observed in other LIAs [19-22]. In DARHT Axis-II this interaction causes a slow beam sweep, which is corrected by application of a steering dipole fields at a few locations in the LIA [5].

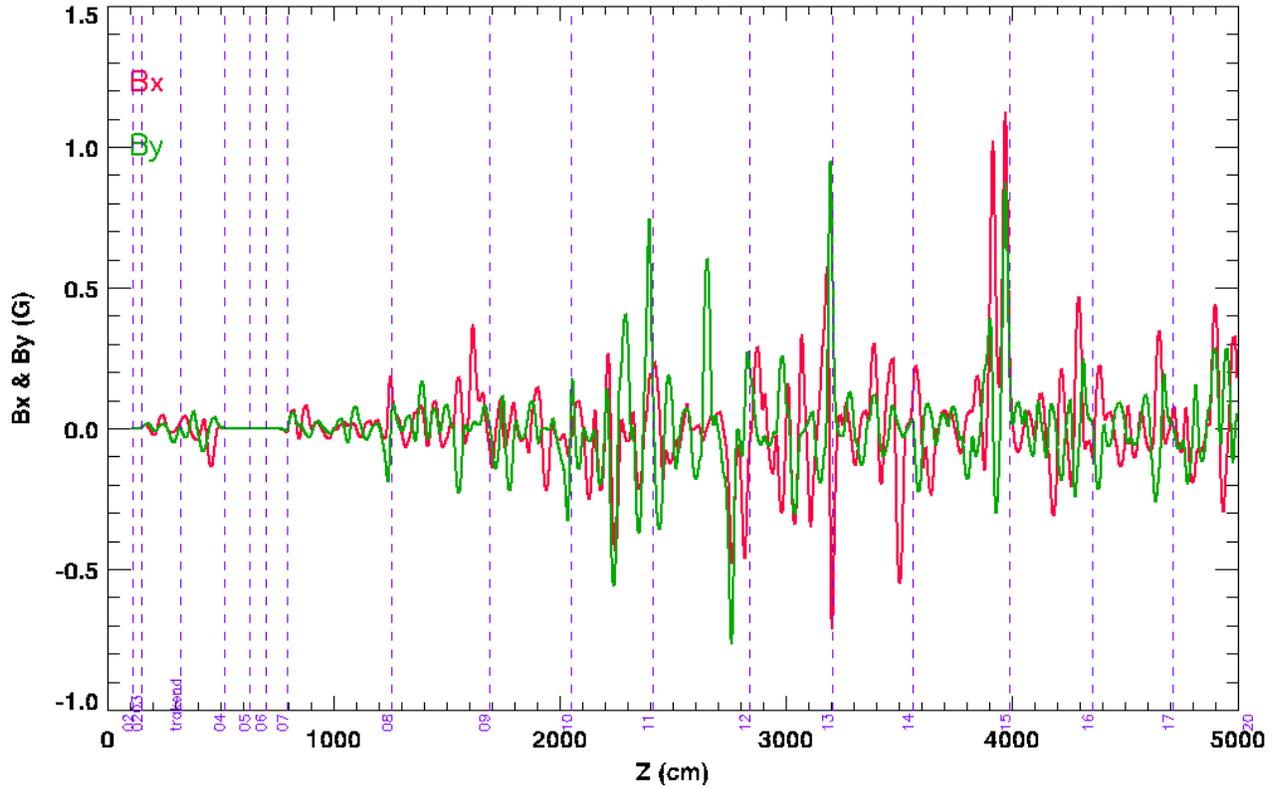

Figure 8: Transverse magnetic field on axis from measured cell misalignment.

The transverse magnetic fields on axis resulting from only the cell misalignments are shown in Fig. 8 for the tune in Fig. 1. These fields were used in an LSP-Slice run to assess the effect of cell misalignment on beam emittance in the DARHT Axis-II LIA. As seen in Fig. 9, misalignment fields have essentially no effect on the beam emittance. Moreover, even increasing the misalignment fields by an order of magnitude would only produce a ~20% emittance growth.

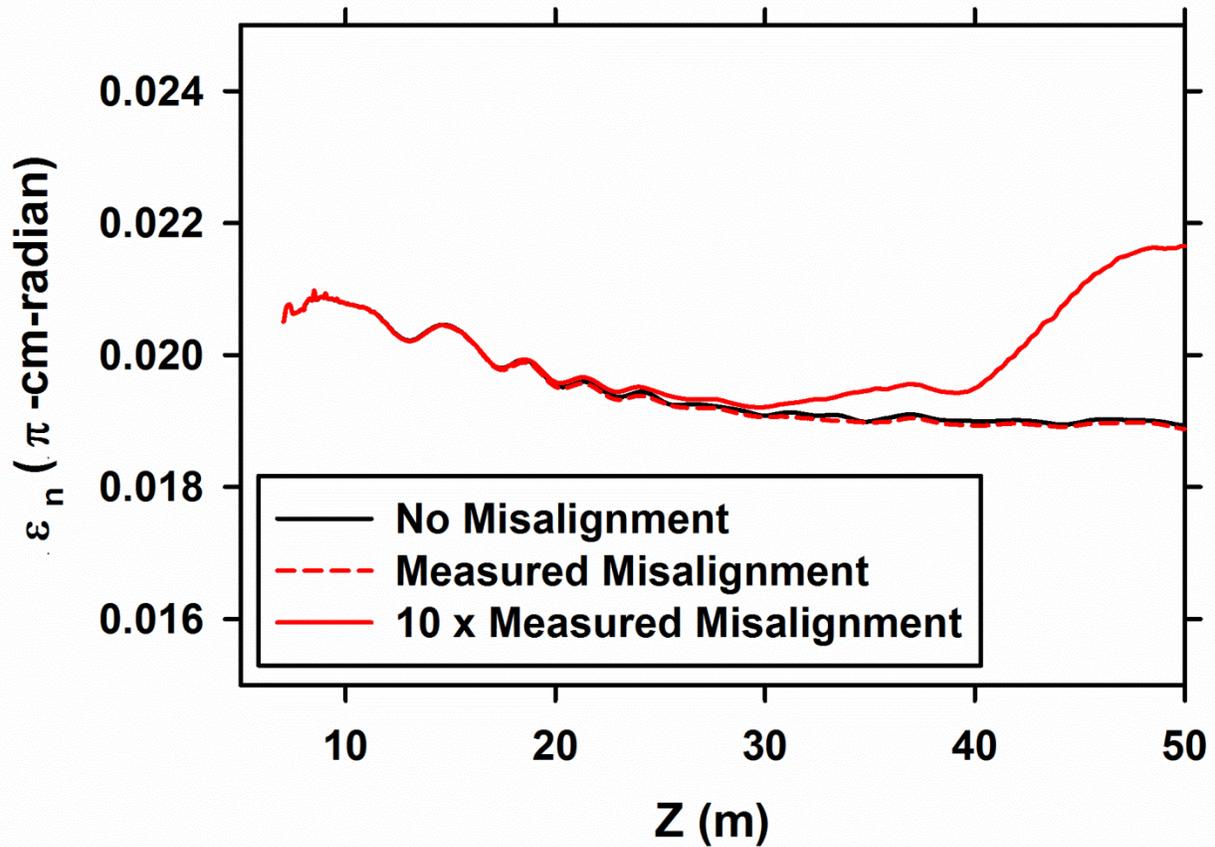

Figure 9: Normalized emittance growth caused by transverse fields due to misalignment of cells.

In practice we apply steering dipole fields to correct the beam sweep caused by misalignment fields. Since the transverse fields from the steering coils are an order of magnitude greater than the misalignment fields plotted in Fig. 9 their effect on emittance must be explored. To do this using LSP-Slice, I used the steering dipole fields from an XTR simulation. These were the actual steering fields that we used on the LIA throughout most of 2013. The steering fields for these currents are shown in Fig. 10, along with the misalignment fields. The emittance growth caused by the combined misalignment and steering fields is quite small. As shown in Fig. 11 the combined effect of these transverse fields is only a 2.6% emittance increase in the LIA. Therefore, we must look elsewhere for the cause of the discrepancy between theory and observations.

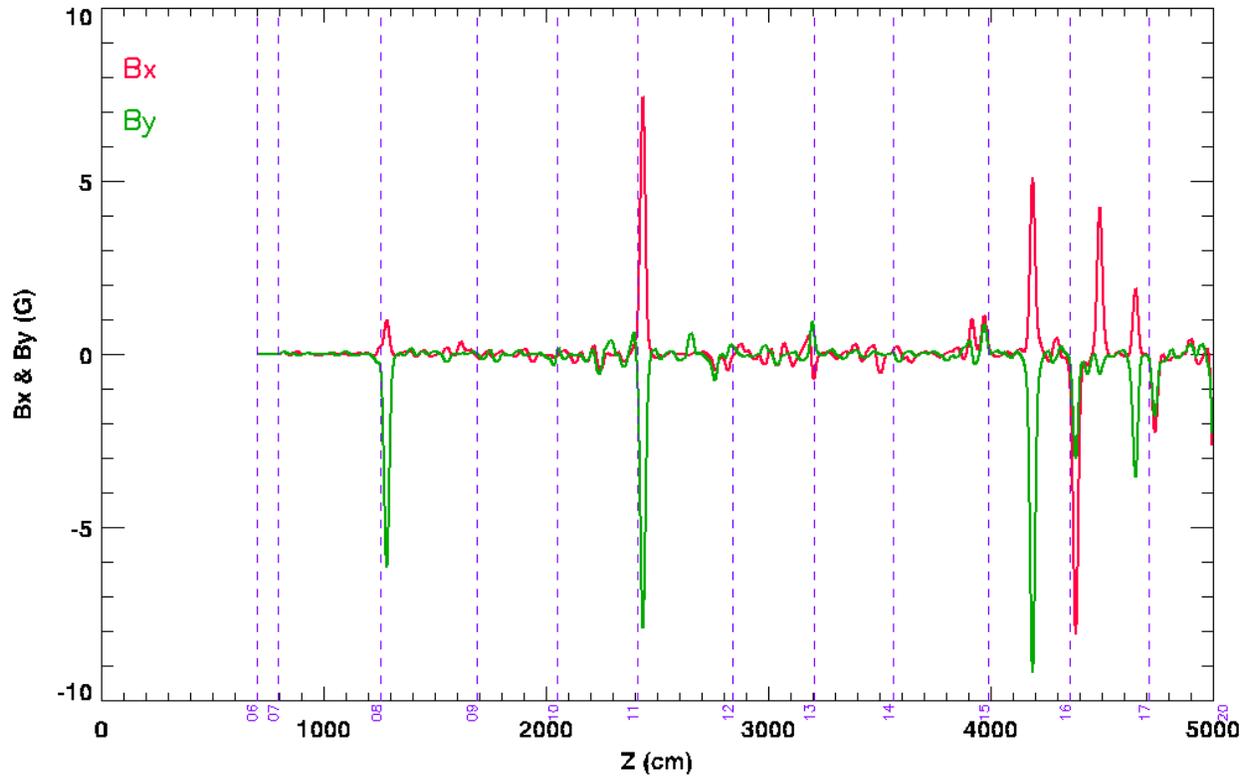

Figure 10: Transverse magnetic fields on axis from steering dipoles and cell misalignments. Note order of magnitude change of scale from Fig. 7.

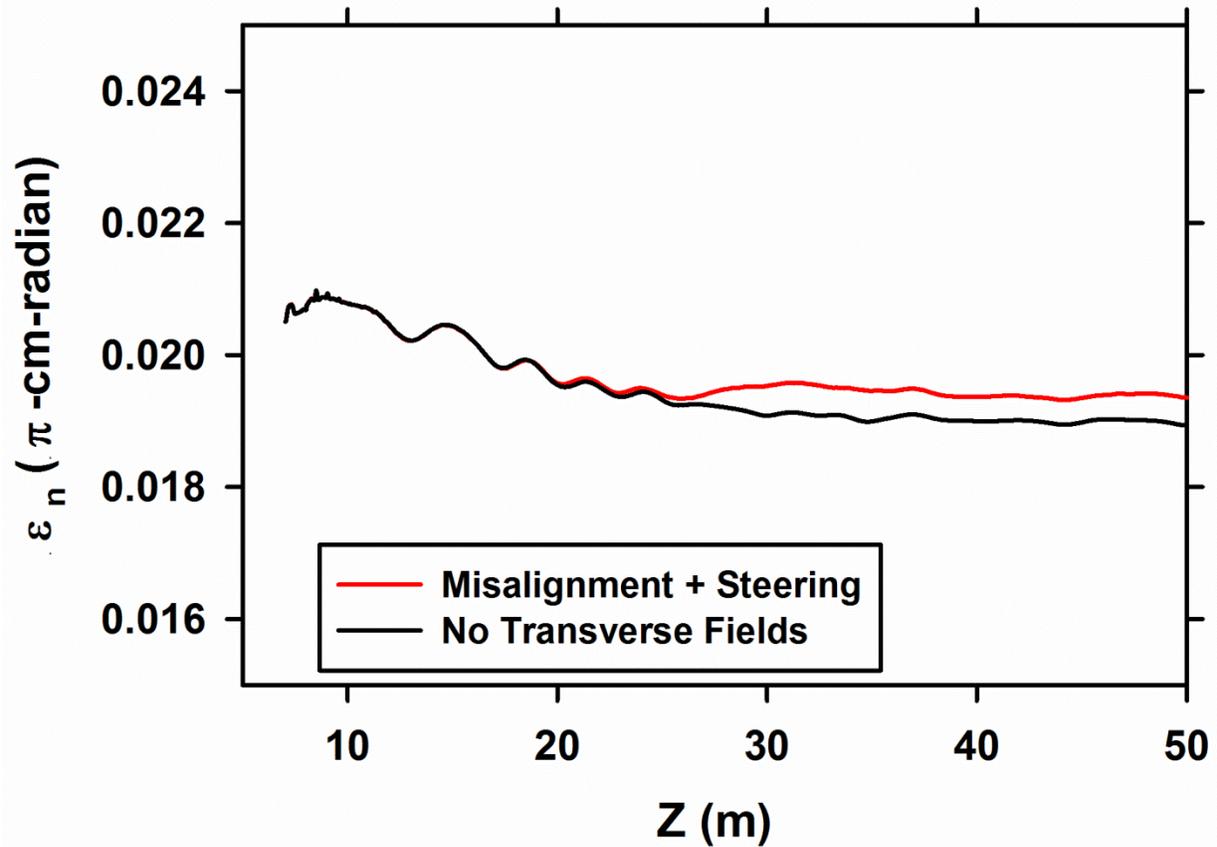

Fig. 11: Normalized emittance growth due to transverse magnetic fields from the combined effects of misalignments and applied steering dipoles.

*Emittance Growth from Envelope Oscillations*

Emittance growth can result from envelope oscillations caused by a mismatch of the beam to the magnetic transport system. A badly mismatched beam exhibits large envelope oscillations, sometimes called a "sausage," "m=0," or "breathing" mode. The free energy in these oscillations feeds the growth of emittance [23]. The detailed mechanism of this contribution to emittance growth is parametric amplification of electron orbits that resonate with the envelope oscillation, expelling those electrons from the beam core into a halo [24,25] (see Appendix A).

Figure 12 shows the envelope oscillations of a mismatched beam and the resulting emittance growth from an earlier simulation [26]. Halo growth was quite clear in LSP-slice movies of the beam distribution as it propagated through the LIA. Several frames of the movie are displayed in Fig. 13 to illustrate the evolution of the halo in configuration and phase space.

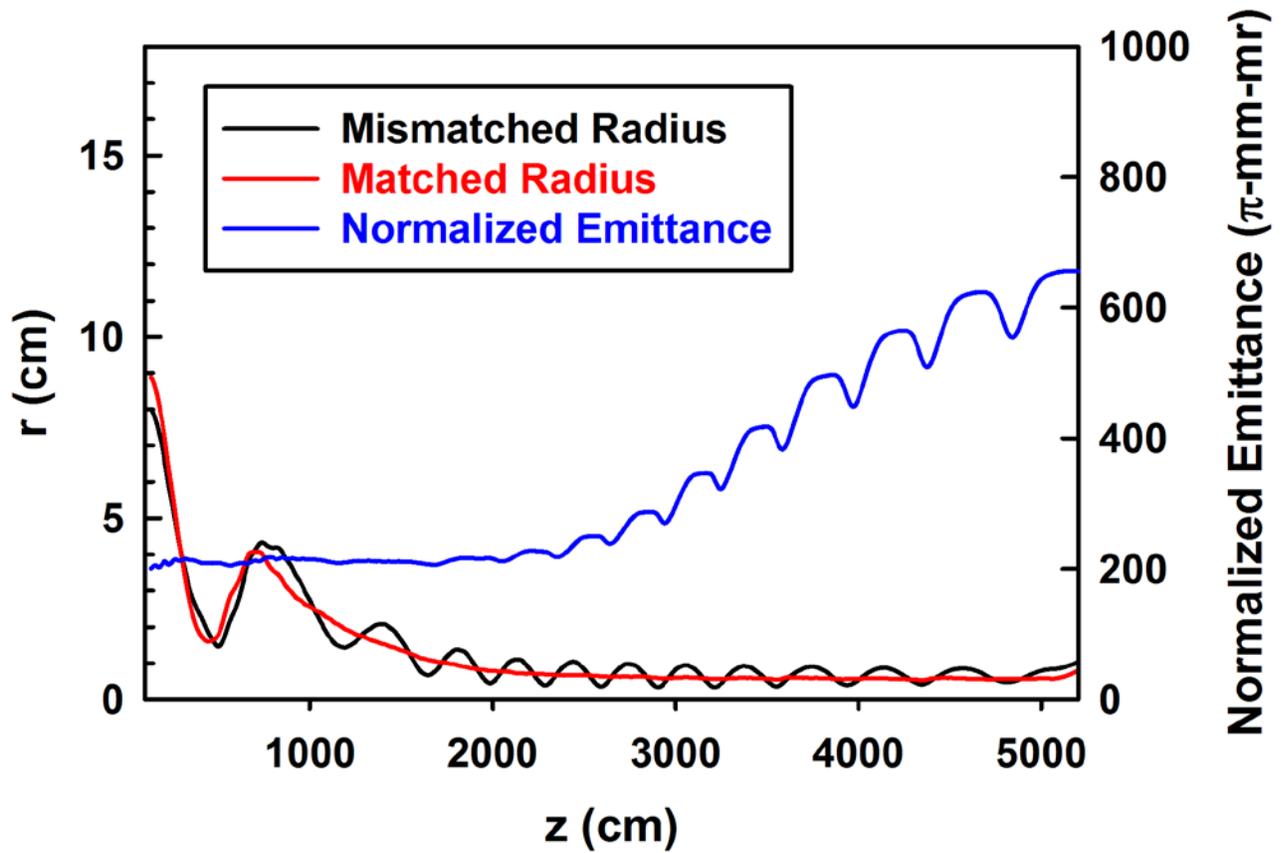

Figure 12: (Black) Envelope radius of the mismatched beam. (Red) Envelope radius of a matched beam. (Blue) The normalized emittance of the mismatched beam simulated by the 1D version of LSP-slice, showing substantial growth through the LIA. (1000 π-mm-mr=0.10 π-cm-radian) (*Adapted from ref. 26*)

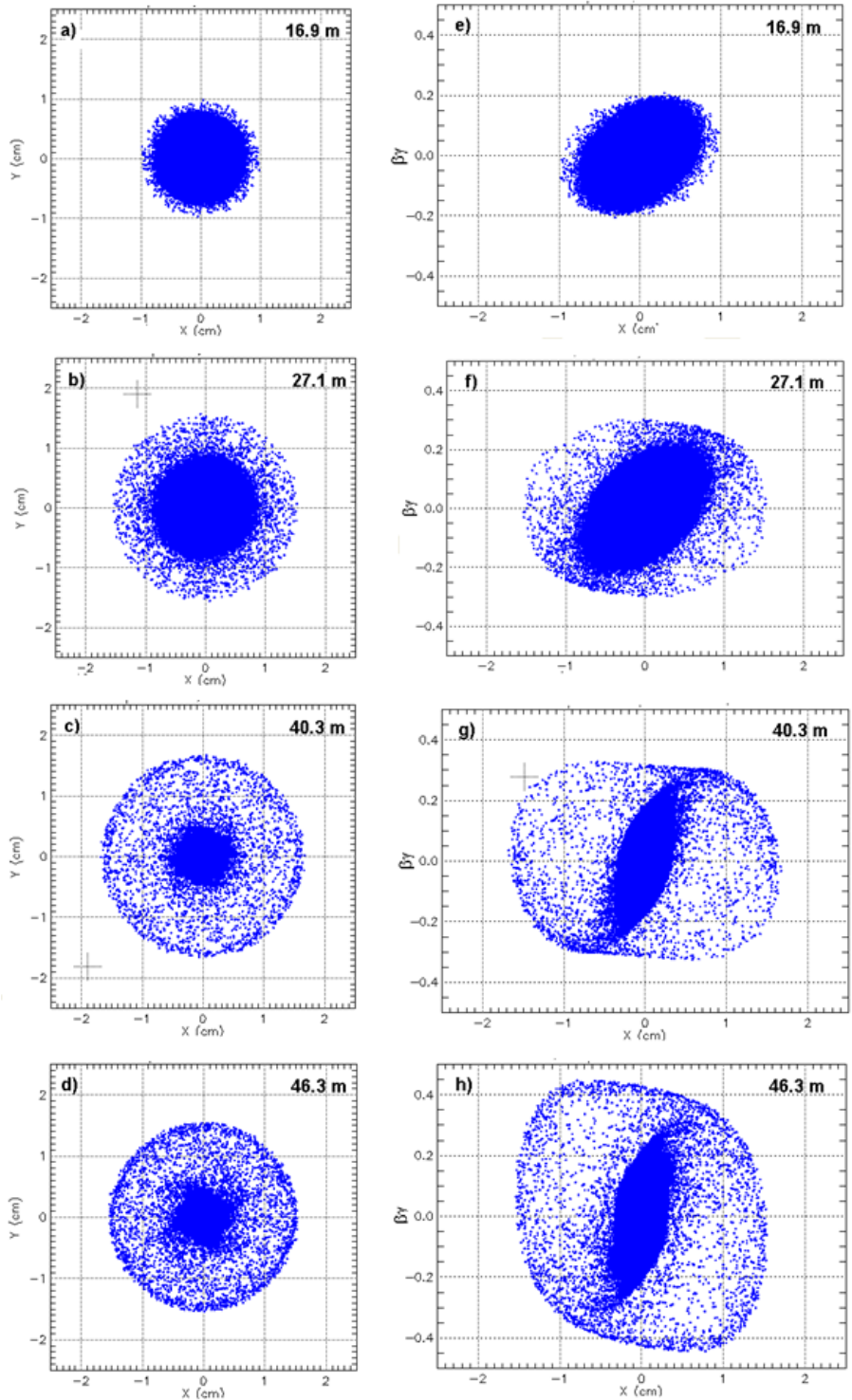

Figure 13: Left column (a, b, c, d): Configuration space ($x, y$) showing growth of halo as mismatched beam propagates through LIA (top to bottom). Right column (d, e, f, g): Phase space ($x, p_x$) showing the increase in phase area (proportional to emittance) as resonant electrons are ejected into the halo. (*Adapted from ref. 26*)

The slice in this earlier simulation was launched at the diode exit in order to demonstrate the lack of emittance growth in the injector and BCUZ, even though the beam envelope has large amplitude excursions there. In order to directly compare the effect of envelope oscillations with the other effects discussed above, I performed a new series of mismatch simulations that were launched at the same point shown in Fig. 1. Mismatch can be accomplished by varying initial slice parameters from the nominal values given in Table I. Since the initial slice should be launched with $r_0' = 0$, I explored varying the initial radius and the wall potential. Figure 14 compares the sensitivity of emittance growth from these perturbations in the initial slice parameters. Apparently, they are equally effective for producing the mismatch and emittance growth. However, varying $r_0$ also affects the beam rotation because it defines the flux required to zero the canonical momentum; $P_\theta = \gamma_0 m_e r_0^2 \omega - e r_0 A_{\theta 0} = 0$, where $A_{\theta 0} \approx r_0 B_z / 2$. Therefore, I only used variations in the initial wall potential $\varphi_{Wall}$ to excite the envelope oscillations for this investigation.

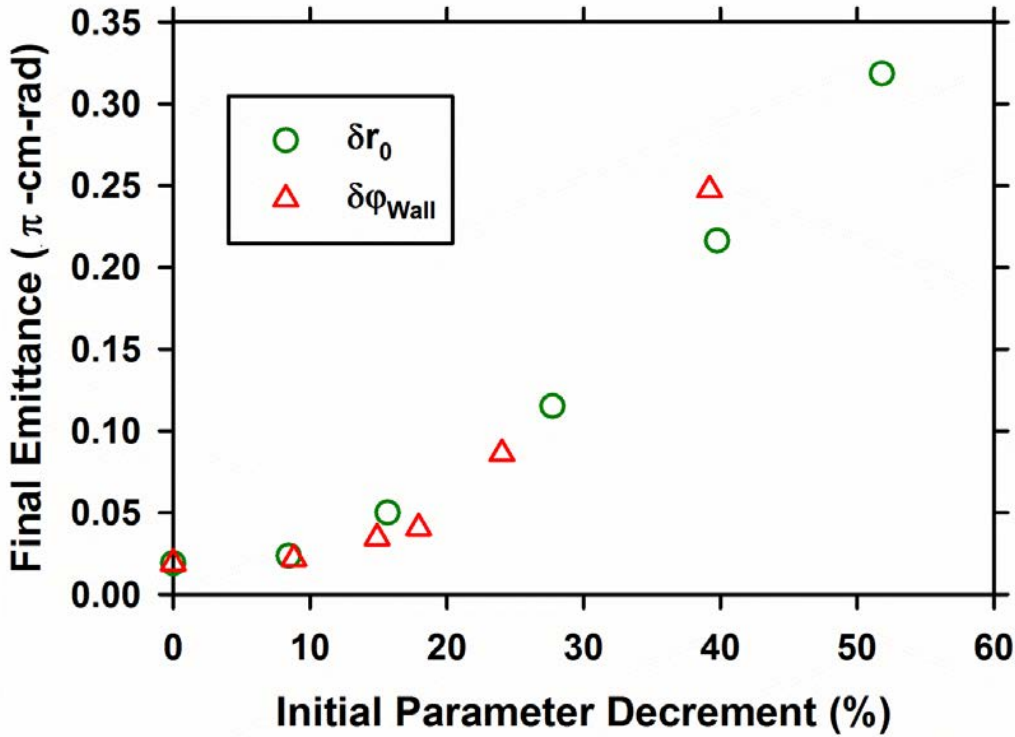

Figure 14: Comparison of causes of beam mismatch. The final emittance at $z = 52$ m is plotted vs the percentage decrement of the initial radius ($\delta r_0$, green circles) or wall potential ($\delta \Phi_{Wall}$, red triangles) from their nominal values in Table 1. The 1D version of the code was used for this analysis.

Several striking features of this mechanism are evident from the simulation results.

- There is a threshold of oscillation amplitude for emittance growth.
- When the initial envelope oscillations are small, the emittance grows almost linearly
- When the initial envelope oscillations are large, the emittance rapidly grows and then saturates.
- The large halo generated on these severely mismatched beams appears to damp the oscillations after the emittance saturates.
- The most severe cases show evidence of multiple halos.

Earlier investigations used a relatively crude metric of envelope oscillation amplitude based on relative maxima and minima, which can be effected by observer bias. For this investigation the amplitude was more objectively defined using the full 2D-Slice data record during the linear phase of the growth. The envelope oscillations were quantified by the amplitude normalized to the size of the matched envelope $\eta = (r - r_m)/r_m$, where $r$ is the mismatched beam envelope and $r_m$ is the matched beam envelope at the same position. Then, the rms value $\eta_{rms}$ was calculated over four complete cycles of envelope oscillations near the start of observable emittance growth. The final emittance ~1.5 m past the LIA exit at $z = 52$ m is plotted in Fig. 15 as a function of $\eta_{rms}$. The onset of emittance growth is evident; with $\eta_{rms}$ less than ~0.13 the growth is probably not measurable.

The definition of $\eta_{rms}$ enables a useful means for predicting emittance growth from envelope code calculations. It can be readily calculated from the output of any simulation that produces envelope oscillations, and then used with Fig. 15 to predict the emittance growth that would appear in a full PIC simulation or experiment. For example, XTR simulations of the tune shown in Fig. 1 indicate that increasing the excitation current of the first three solenoids in the second cell block causes noticeable envelope oscillations with $\eta_{rms} = 0.1898$. Inspection of Fig. 16 suggests that the resulting emittance would only be slightly increased to ~0.0230 π-cm-rad.

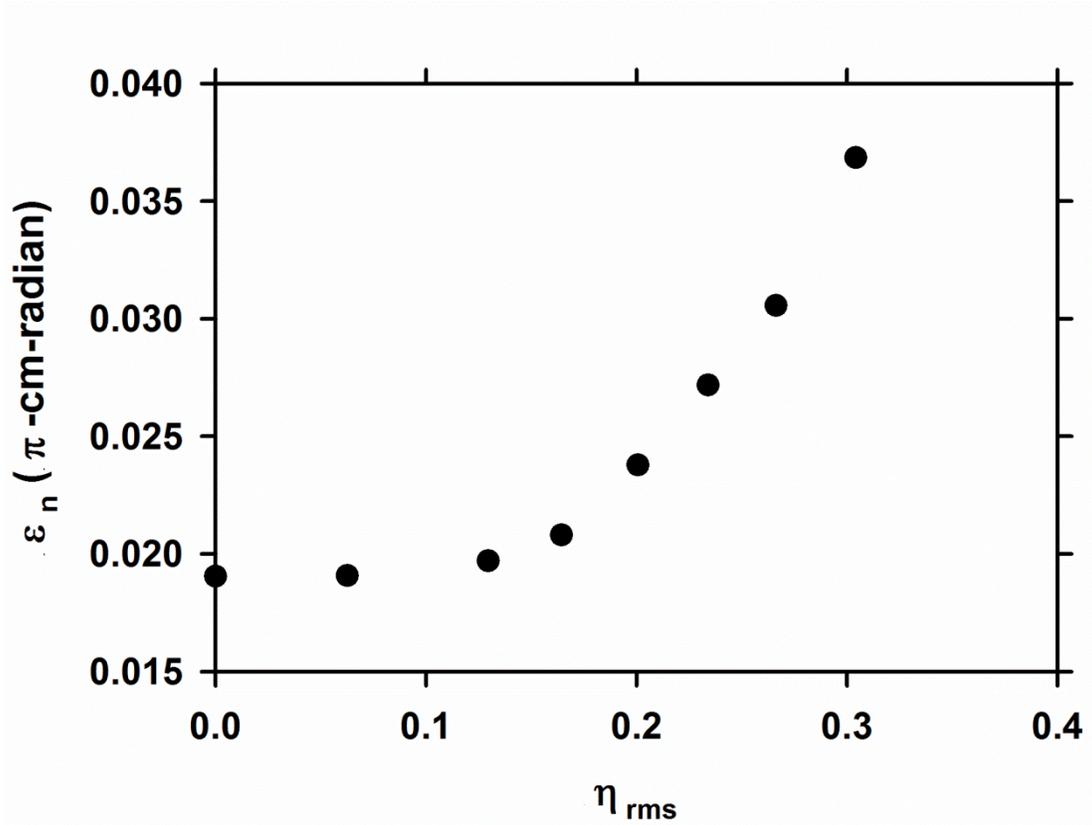

Figure 15. Final normalized emittance at $z = 52$ m as a function of normalized oscillation amplitude.

Linear emittance growth on a weakly mismatched beam is illustrated in Fig. 16. Weakly mismatched beam halos are diffuse, becoming more distinct for worse mismatch and larger envelope oscillations. Rapid emittance growth and saturation on a severely mismatched beam is shown in Fig. 17. Strongly excited halos have one or more distinct rings. The halos of the most severely mismatched beams develop multiple distinct rings, corresponding to higher order resonances. Figure 18 compares the configuration space distribution just outside the LIA exit for several representative cases of emittance growth, clearly showing multiple rings in the halos of the most severe cases, and diffuse halos for the weak cases. The possibility that the distinct rings in the largest halos result from edge focusing by spherical aberration is yet to be investigated.

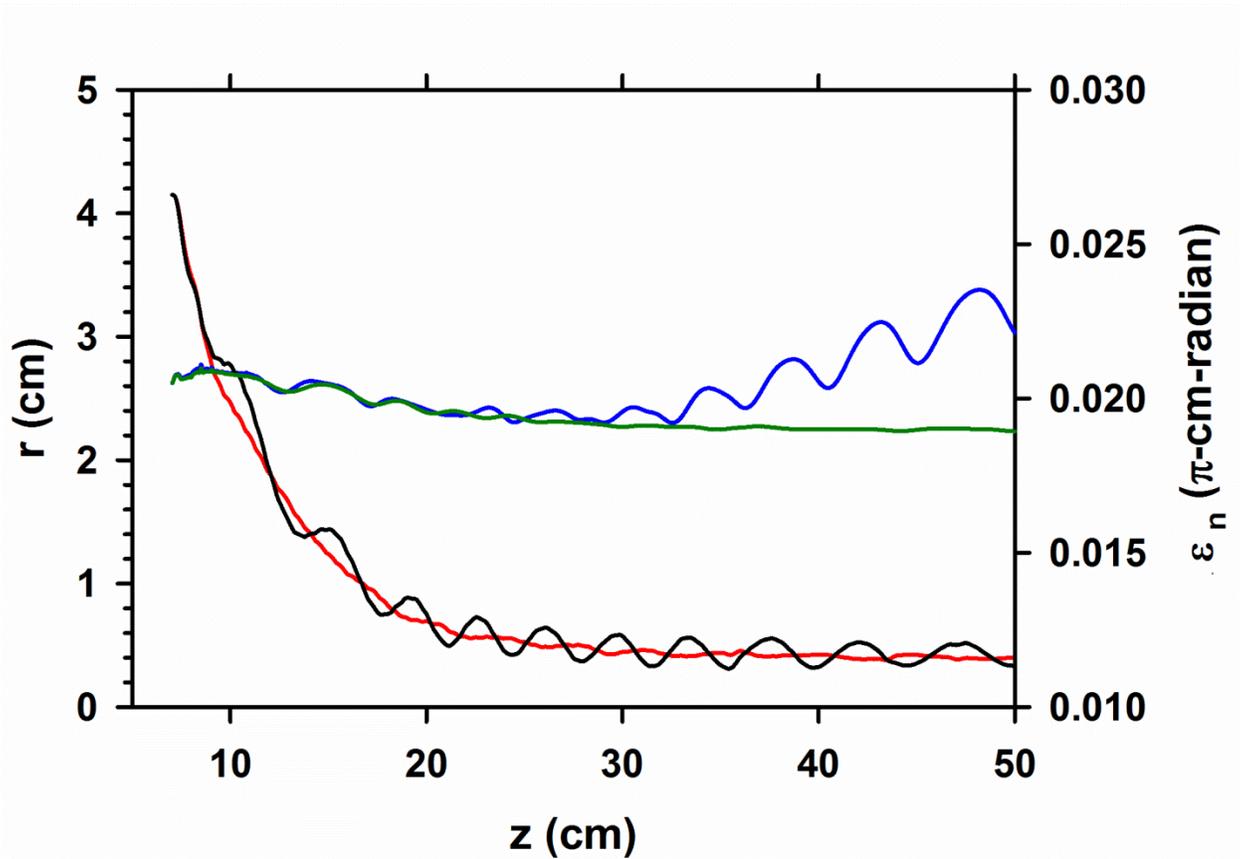

Figure 16: (Black) Envelope radius of a weakly mismatched beam simulated by the 2D version of LSP-slice. The envelope oscillation normalized amplitude in this simulation is $\eta_{rms} = 0.2$. (Red) Envelope radius of a matched beam. (Green) The normalized emittance of the matched beam. (Blue) The normalized emittance of the mismatched beam, showing linear growth through the LIA.

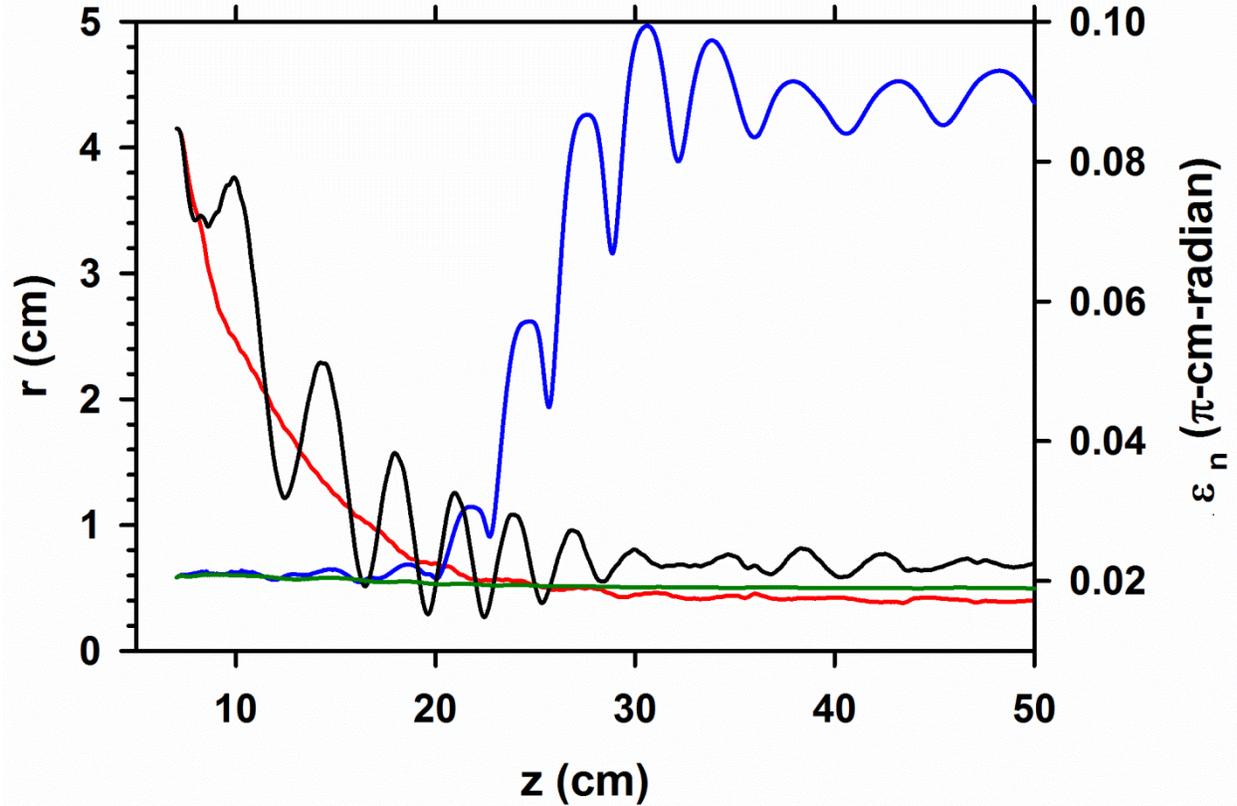

Figure 17: (Black) Envelope radius of a severely mismatched beam showing damping of the oscillations. The envelope oscillation normalized amplitude in this simulation is $\eta_{rms} = 0.5$. (Red) Envelope radius of a matched beam. (Green) The normalized emittance of the matched beam. (Blue) The normalized emittance of the mismatched beam, showing rapid growth and saturation. (Note the emittance scale difference from Fig. 16).

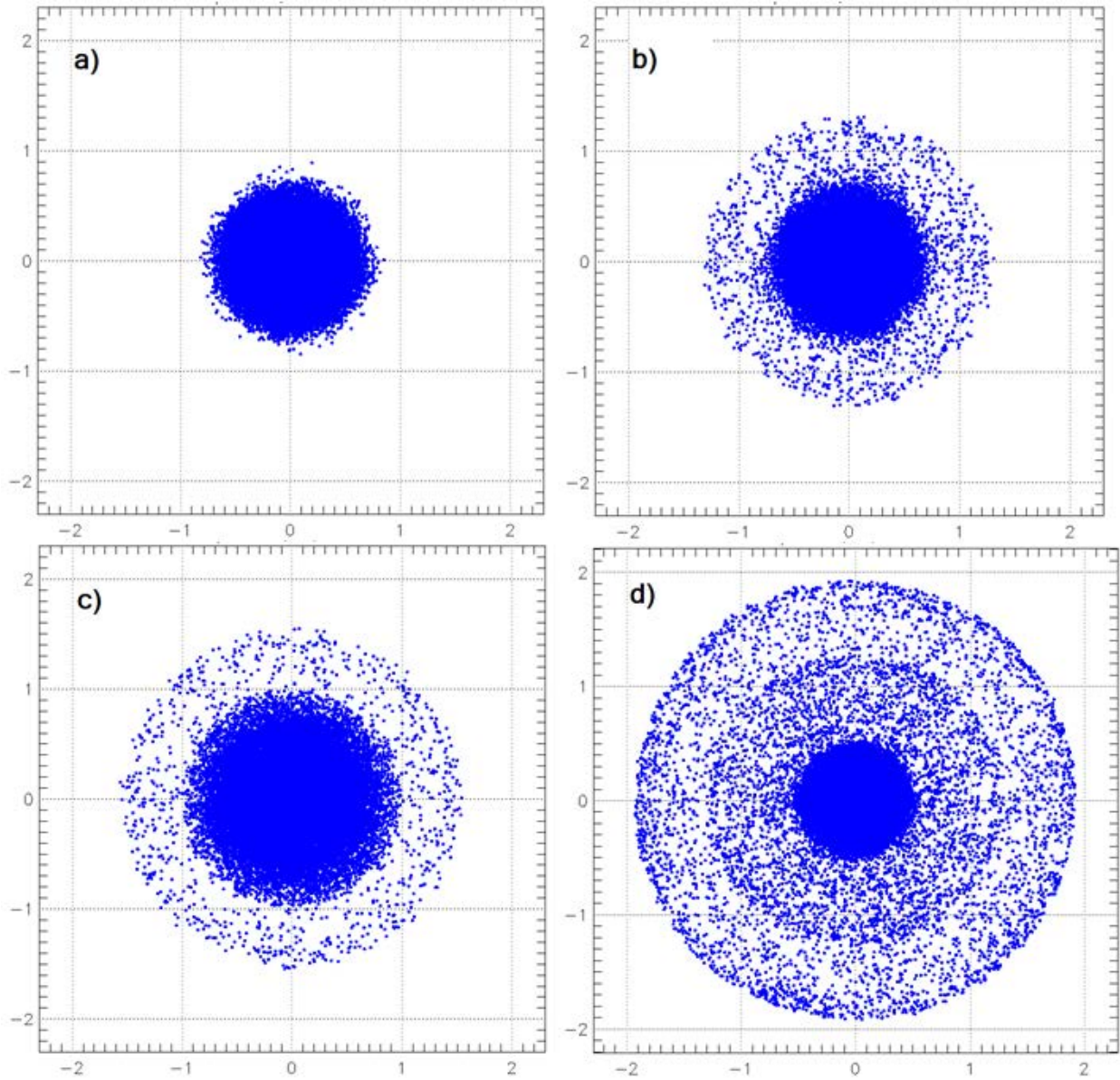

Figure 18: Configuration space plots of the beam distribution at $z = 52$ m, which is ~1.5 m past the LIA exit. Dimensions are in cm. a) The matched case with $\eta_{rms} = 0.0$, $\varepsilon_n = 0.0191$ $\pi$-mm-mr  b) The weakly mismatched case shown in Fig. 16 with $\eta_{rms} = 0.2$, $\varepsilon_n = 0.0238$ $\pi$-mm-mr  c) $\eta_{rms} = 0.3$, $\varepsilon_n = 0.0369$ $\pi$-mm-mr  d) The severely mismatched case shown in Fig. 17 with $\eta_{rms} = 0.5$, $\varepsilon_n = 0.0901$ $\pi$-mm-mr .

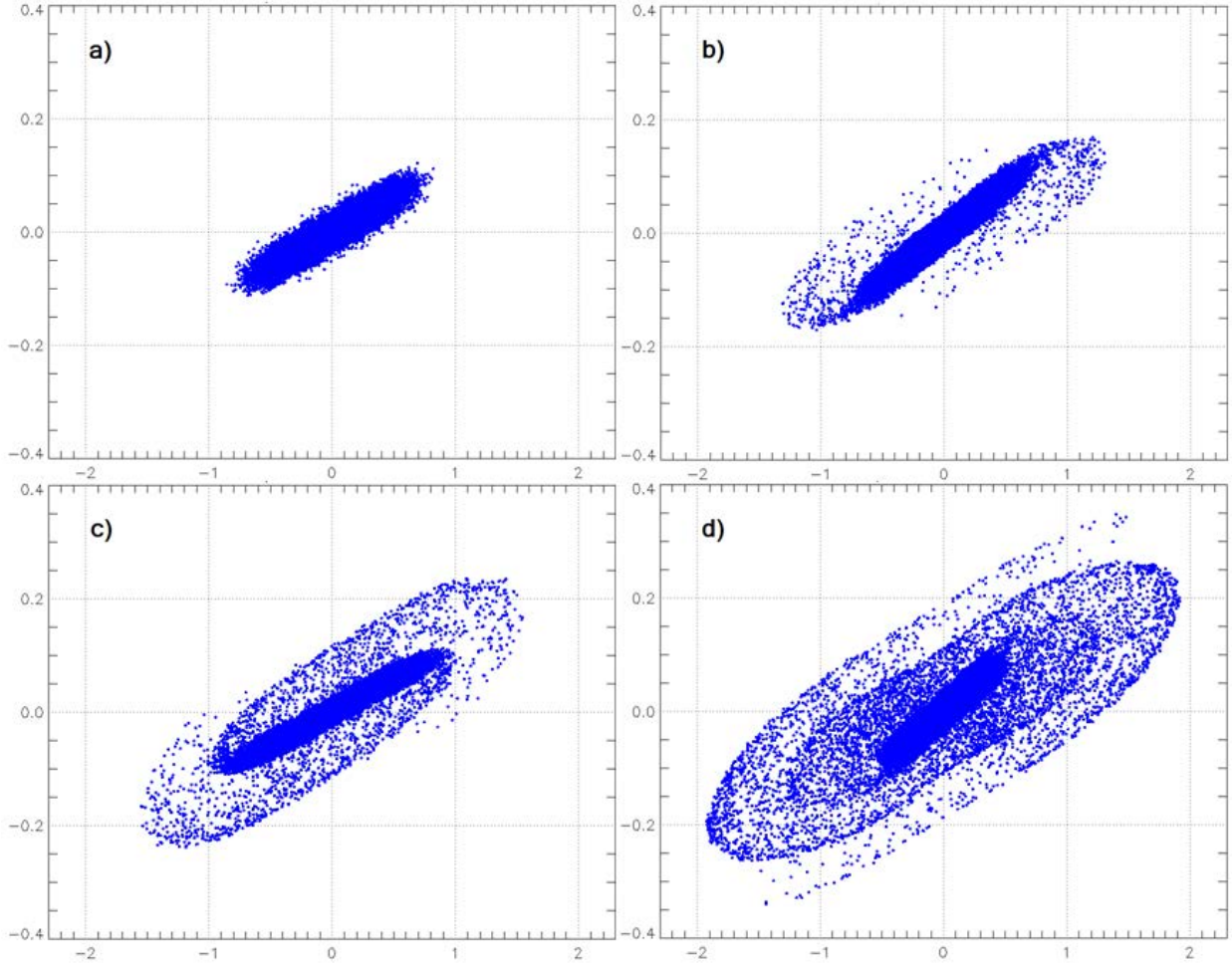

Figure 19: Phase space plots of the beam distribution at $z = 52$ m, which is ~1.5 m past the LIA exit. Plots correspond to the configuration space plots in Fig. 18. Spatial dimensions are cm and momentum dimensions are transverse $\beta\gamma$. a) The matched case with $\eta_{rms} = 0.0$, $\varepsilon_n = 0.0191$ $\pi$-mm-mr b) The weakly mismatched case shown in Fig. 16 with $\eta_{rms} = 0.2$, $\varepsilon_n = 0.0238$ $\pi$-mm-mr c) $\eta_{rms} = 0.3$, $\varepsilon_n = 0.0369$ $\pi$-mm-mr d) The severely mismatched case shown in Fig. 17 with $\eta_{rms} = 0.5$, $\varepsilon_n = 0.0901$ $\pi$-mm-mr.

## III. CONCLUSIONS

The Axis-II beam is centered to less 0.5-cm of the axis at the BPMs throughout most the accelerator, so it is doubtful that there is much emittance growth caused by large gyro-radius effects. From these simulations, we estimate growth from this cause to be no more than ~10%, and likely much less.

Emittance growth from the misalignment dipole fields was shown to be insignificant, with a tenfold increase necessary to have a noticeable emittance growth. Moreover, although we apply a number of steering dipoles to correct for beam motion, these simulations show that growth due to steering dipoles in addition to the misalignments is insignificant; < 3% at most.

If growth in the LIA is indeed responsible for the final ~1000 π-mm-mr, suggested by early measurements, then it is most likely due to envelope oscillations resulting from beam mismatch. Halo growth from the parametric amplification of orbits by the envelope oscillations significantly increased the emittance in these simulations. However, growth to ~1000 π-mm-mr would only result from a very severe mismatch, with normalized oscillation amplitude ~ 50% or more. This severe a mismatch would indicate that our diode simulations of initial conditions are grossly inaccurate.

Improving the beam match to reduce envelope oscillations would reduce emittance growth from this cause, and would thereby improve radiographic resolution. The design of our tunes features the ability to improve the match by varying only the first few solenoids after the BCUZ [26]. Therefore, we plan to use this feature in an attempt to reduce the emittance of the DARHT Axis-II beam, in order to improve the radiographic source spot.

As mentioned in the introduction, we will also investigate the possibility that the higher than expected emittance is due to imperfections in the diode, such as non-uniform cathode emission, or inaccurate positioning, which are not accounted for in our simulations to date.

Acknowledgements

The author appreciates stimulating discussions with his colleagues on this topic, and many others. Especially noteworthy are technical conversations with D. Moir and M. Schulze about emittance and its measurement at DARHT. Help with the LSP-Slice code and its multiprocessor implementation from Carsten Thoma and Chris Mostrum at Voss Scientific is gratefully acknowledged.

This research was supported by the U. S. Department of Energy and the National Nuclear Security Administration under contract DE-AC52-06NA25396.

Appendix A

Following Gluckstern [24] we consider a uniform axisymmetric beam with envelope radius $r_{env}$ in a constant external focusing field. In Cartesian coordinates the equation of motion for an electron inside the envelope is

$$x'' + k_0^2 x = \frac{K}{r_{env}^2} \quad (1.7)$$

with the same equation for $y$. Here, primes denote differentiation with respect to $z$, $K$ is the generalized perveance described in Section I, and $k_0$ is the wavenumber of oscillations with no space-charge. Assume a mild breathing mode oscillation of the beam envelope the beam envelope, $r_{env} \to r_0(1 - \delta \cos pz)$ and expand in the small parameter $\delta \ll 1$. For electrons inside the envelope Eq. (1.1) becomes

$$x'' + k^2 x = \frac{2\delta K}{r_0^2} x \cos pz \quad (1.8)$$

With $k^2 = k_0^2 - \kappa / r_0^2$ the tune depressed wavenumber. Multiplying by $4/p^2$ casts this equation into the well-known Mathieu equation, which admits both stable and unstable solutions;

$$\frac{d^2 x}{d\zeta^2} + (a - 2q \cos 2\zeta) x = 0 \quad (1.9)$$

where $\zeta = pz/2$, and

$$a = \frac{4}{p^2}\left(k_0^2 - K/r_0^2\right)$$
$$q = \frac{8K\delta}{p^2 r_0^2} \quad (1.10)$$

The Mathieu equation has been extensively studied, because of its many applications to physics and engineering. Boundaries of stable regions are tabulated as characteristic functions $a_n(q)$. Figure A1 shows the zones of stability bounded by these characteristic functions. Electrons with $a, q$ falling in an unstable region are ejected from the core of the beam into the halo.

From Fig. A1, it is evident that for $a < q$ the solutions are unstable almost everywhere. The exception to the $a < q$ rule-of-thumb criterion for instability is near the origin for $a, q < 1$. This region of the stability diagram is blown up in Fig. A2, showing the approximate stability boundary $a \sim 1 - q - q^2/8$. Thus, $a > 1 - q$ is unstable when $a < 1$.

For $a \gg q$, there are large zones of stability. For beam parameters of interest, and small envelope oscillations, $q \ll 1$. Near this axis, orbits with $a = n^2, n = 1, 2, 3\ldots$ are unstable. That is, orbits are unstable if they are harmonics of the envelope modulation. The lowest order harmonic, $k = 2p$ for $n = 1$ is the one most discussed in halo formation theoretical articles.

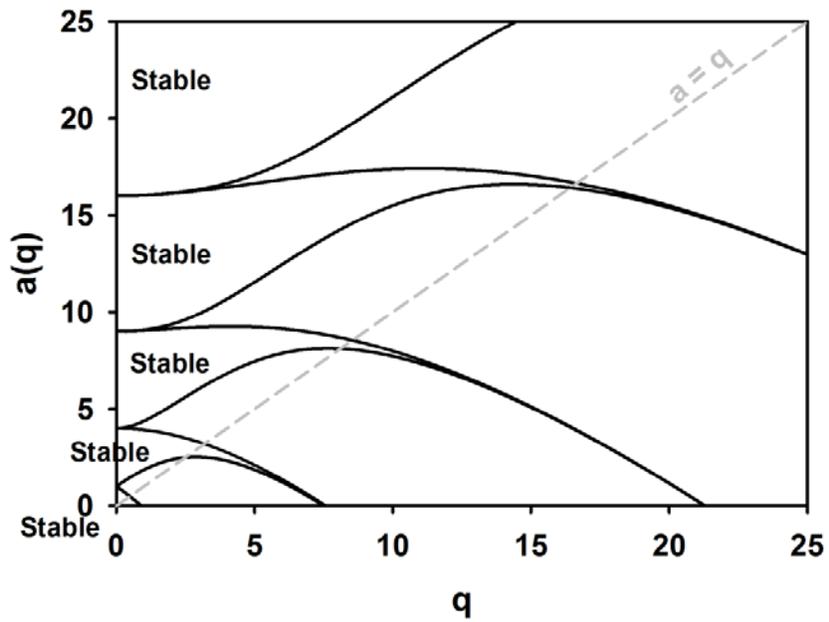

Figure 1: Stable and unstable regions of solutions to the Mathieu equation

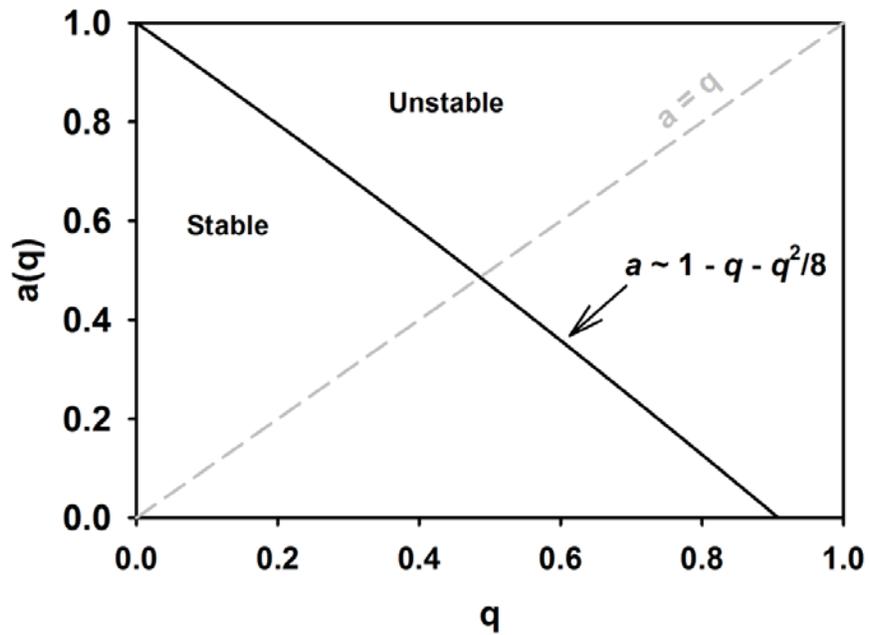

Figure A2: Expanded view of Fig. 1 showing stable and unstable regions near the origin.